\documentclass[%
 reprint, 
 amsmath,amssymb,
 aps,
prx,
]{revtex4-2}
\usepackage{savesym}
\savesymbol{corresponds} 

\usepackage{graphicx}
\usepackage{dcolumn}
\usepackage{bm}
\usepackage{booktabs}
\usepackage{xcolor}
\usepackage{braket}
\usepackage{comment}
\usepackage{makecell}
\usepackage{mathabx}  
\usepackage[version=4]{mhchem}

\DeclareMathOperator{\Tr}{Tr} 

\begin{document}

\preprint{APS/123-QED}

\title{How NOT to build control-target gates in semiconductor quantum dots and beyond}

\author{Roman Korol}
\email{roman.korol@sherbrooke.ca}
\affiliation{Department of Chemistry, University of Rochester, Rochester, New York 14627, United States}
\affiliation{Département de Chimie, Université de Sherbrooke, Sherbrooke, Québec J1K 2R1, Canada}
\author{John Nichol}%
\email{john.nichol@rochester.edu}
\affiliation{Department of Physics and Astronomy, University of Rochester, Rochester, New York 14627, United States}%

\author{Ignacio Franco}
\email{ignacio.franco@rochester.edu}
\affiliation{
    Department of Chemistry, University of Rochester, Rochester, NY, USA}
\affiliation{
    Department of Physics and Astronomy, University of Rochester, Rochester, NY, USA}
\affiliation{
    The Institute of Optics, University of Rochester, Rochester, NY, USA}

\date{\today}

\begin{abstract}
Universal quantum computation requires single-qubit control together with at least one entangling two-qubit gate. CNOT and CROT gates are two famous examples of such gates, wherein the state of one qubit (control) dictates the transformation applied to the other (target). By using a simple derivation motivated by symmetry, we show that current device architectures of semiconductor-based quantum dot devices prevent efficient implementation of a CNOT and other asymmetric control-target gates via Heisenberg exchange, Coulomb repulsion, or other interaction that is invariant under spin exchange. Guided by this general principle, we propose a heterogeneous blueprint of double quantum dot devices that enables efficient implementation of the CNOT by breaking the spin exchange symmetry. Crucially, our numerical simulations predict that this novel device blueprint can enable a single-pulse fault-tolerant $100$~ns CNOT gate in isotopically purified silicon.
\end{abstract}

\maketitle

\section{\label{sec:intro}Introduction}

Universal quantum computation hinges on the ability to execute a set of single- and two-qubit quantum gates with high fidelity across a scalable qubit architecture~\cite{nielsenQuantumComputationQuantum2011}. Single qubit gate fidelities exceeding the fault tolerance mark have been demonstrated across various leading quantum computing platforms: semiconductor spin qubits surpassing four nines ($>99.99\%$)~\cite{lawrieSimultaneousSinglequbitDriving2023}, superconducting circuits above six nines~\cite{liErrorSinglequbitGate2023}, and trapped ions approaching nine nines~\cite{smithSinglequbitGatesErrors2025}. In contrast, for two-qubit gates the gate fidelity lags behind with recently reported fidelity of $99.99\%$ for ion qubits -- record-high across all platforms ~\cite{hughesTrappedionTwoqubitGates2025} -- at the cost of relatively slow operation times ($\sim200$~$\mu $s). Semiconductor spins~\cite{noiriFastUniversalQuantum2022} and superconducting circuits~\cite{dingHighfidelityFrequencyflexibleTwoqubit2023,liRealizationHighfidelityCZ2024,marxer999FidelitySinglequbit2025} provide faster but less precise control with the highest two-qubit gate fidelity reported at the level of three nines. Why are the two-qubit gates more prone to error, and, more importantly, how can we mitigate that?

Many two-qubit gates require not one but a series of operations, sometimes including more than one two-qubit interaction~\cite{liControllableExchangeCoupling2012,klinovajaExchangebasedCNOTGates2012,zajacResonantlyDrivenCNOT2018,veldhorstTwoqubitLogicGate2015}. This requirement increases the total gate time and accumulates errors, contributing to the lower gate fidelity for two-qubit gates compared to the gates achieved in a single shot~\cite{schuchNaturalTwoqubitGate2003}. Thus, designing simpler two-qubit operation sequences is of great interest. Specifically, we show that additional operations are unavoidable in devices that use a single qubit blueprint if the interaction between qubits is symmetric, while the target gate is not. Both Coulomb repulsion and Heisenberg exchange between identical particles are symmetric with respect to particle exchange. The Heisenberg exchange interaction is used to couple semiconductor quantum dot qubits~\cite{burkardSemiconductorSpinQubits2023} and Coulomb repulsion is at the core of capacitive coupling in quantum dots~\cite{taylorFaulttolerantArchitectureQuantum2005}, superconducting circuits~\cite{rasmussenSuperconductingCircuitCompanion2021}, and ion qubits~\cite{sorensenEntanglementQuantumComputation2000}.  Because scalability of the device is typically achieved by repeating a single-qubit building block, a homogeneous array of identically constructed qubits is generated, and the symmetry of both Coulomb and Heisenberg interactions is preserved.
This, as we show here, makes it impossible to \textit{directly} implement any asymmetric gate, such as controlled-not (CNOT) or controlled-rotation (CROT). To enable the implementation of these asymmetric gates, the exchange symmetry is broken by other means, for instance, via spin-orbit interaction in hole-spin qubits~\cite{geyerAnisotropicExchangeInteraction2024} or by selectively driving one of the two QD qubits~\cite{zajacResonantlyDrivenCNOT2018}. As a last resort, exchange symmetry is broken by sandwiching the symmetric interaction with additional qubit operations~\cite{liControllableExchangeCoupling2012, klinovajaExchangebasedCNOTGates2012}, leading to longer gate times and larger errors.

Here we articulate the mismatch in symmetry between a symmetric two-qubit interaction and an asymmetric target gate as a core reason why additional interactions are needed. Based on this analysis, we suggest a novel and general approach to overcome it via a heterogeneous device architecture. Specifically, we show that when devices are constructed in a homogeneous fashion by repeating one qubit blueprint throughout, the symmetry of the underlying interqubit interaction is necessarily preserved. In contrast, by defining neighboring qubits in non-identical ways, the effective interqubit interaction in the computational subspace can be asymmetric, facilitating the implementation of asymmetric gates. Although this principle is general, for concreteness, we illustrate it in semiconductor quantum dots. This platform is chosen because of the wide variety of distinct qubit mappings that have been realized on virtually identical hardware. Specifically, we show that by adopting a hybrid qubit encoding that alternates between $ST_0$ and $ST_-$ qubits, a CNOT gate can be realized directly using only the symmetric Heisenberg exchange. Numerical simulations assessing the effects of the main sources of decoherence in these devices -- static noise and leakage out of computational subspace -- yield gate fidelities of over 99\% for experimentally accessible parameter regimes, advancing the platform into the fault-tolerant regime based on the commonly quoted $\sim1\%$ surface-code threshold~\cite{wangSurfaceCodeQuantum2011}.

The paper is outlined as follows. We begin with Sec.~\ref{sec:the}, presenting the general proof of the symmetry preservation in homogeneously mapped devices as well as its implications. Sec.~\ref{sec:back} provides general background of the semiconductor quantum dot setup, and can be skipped by specialists.
 The CNOT gate is constructed in Secs.~\ref{sec:the-cnot}-\ref{sec:ideal} and its performance is tested in Secs.~\ref{sec:leakage}-\ref{sec:Noise}. Before concluding, we discuss the implications of our finding for quantum dots and beyond in Sec.~\ref{sec:Disc}.

\section{\label{sec:the}Symmetry Constraints in Qubit Design}
Is it possible to realize an asymmetric two-qubit gate with a single symmetric interaction? In this section, we show that this task is impossible for a device that is built out of homogeneous blocks. However, it can be achieved by utilizing a heterogeneous qubit encoding. 
\subsection{Homogeneous Architecture Preserves Interaction Symmetry}\label{symmetry_proof}

Firstly, we demonstrate that a two-qubit interaction that is symmetric with respect to qubit exchange cannot by itself implement an asymmetric gate (such as CNOT) if the two qubits are encoded identically. 

For this, let the total Hilbert space $\mathcal{H}$ be the tensor product of two identical subsystems, A and B, each containing a qubit:
$$ \mathcal{H} = \mathcal{H}_\text{A} \otimes \mathcal{H}_\text{B}. $$
We define the exchange operator $S_\text{AB}$ that swaps the quantum states of the two subsystems. Since a basis of $\mathcal{H}$ can be constructed from all product states, the linear operator $S_\text{AB}$ can be defined by its action on a general product state:
\begin{equation}
S_\text{AB} (\ket{\psi_\text{A}} \otimes \ket{\phi_\text{B}}) = (-1)^\eta\ket{\phi_\text{A}} \otimes \ket{\psi_\text{B}}
\end{equation}
where $\eta$ is the number of fermions in each subsystem, which accounts for the exchange statistics of identical particles.

The nature of the interaction between subsystems A and B is such that the interaction Hamiltonian $H_\text{int}$ between two subsystems is symmetric with respect to this exchange, meaning it commutes with the exchange operator:
$$ [H_{\text{int}}, S_\text{AB}] = 0. $$

The qubits are defined on both subsystems A and B in the same way, meaning that the projection operator that defines the qubit subspace is identical, i.e., $P_\text{A}=P_\text{B}=P$ and the total projector is
$$ \mathcal{P} = P \otimes P ,$$
where the first (second) operator acts on the Hilbert space of A (B). 
 The effective Hamiltonian within the projected computational subspace is then given by
\begin{equation}
    H'_{\text{int}} = \mathcal{P} H_{\text{int}} \mathcal{P}.\label{eq:H'int}
 \end{equation}  
 
We now show that effective Hamiltonian~\eqref{eq:H'int} acting on the computational subspace must retain the exchange symmetry of the full Hamiltonian, provided the two qubits are formed by an analogous projection of identical subsystems. 

The key step is to recognize that the projection operator $\mathcal{P}$ is symmetric with respect to the subsystem exchange operator. This is a direct consequence of using the same projector $P$ for both subsystems $A$ and $B$. 
A similarity transformation with $S_\text{AB}$ swaps the operators in the tensor product. Because both operators are the same, the projector is invariant:
$$ S_\text{AB} \mathcal{P} S_\text{AB}^{-1} = S_\text{AB} (P \otimes P) S_\text{AB}^{-1} = P \otimes P = \mathcal{P}, $$
which means that the swap and the projection operators commute:
\begin{equation}\label{eq:PS_commute}
    [\mathcal{P}, S_\text{AB}] = 0.
\end{equation}

The projected Hamiltonian $H'_{\text{int}}$ also commutes with the swap operator:
\begin{equation}
    [H'_{\text{int}}, S_\text{AB}] = (\mathcal{P} H_{\text{int}} \mathcal{P}) S_\text{AB} - S_\text{AB} (\mathcal{P} H_{\text{int}} \mathcal{P})=0,\label{eq:H'S}
\end{equation}
since $S_\text{AB}$ commutes with both $\mathcal{P}$ and $H_\text{int}$.
 Equation~\ref{eq:H'S} signifies that $H'_\text{int}$ (the effective interaction within the computational subspace) must remain symmetric. \emph{Thus, a symmetric interaction between identical subsystems cannot by itself implement any asymmetric operation in the (projected) qubit subspace.}

\subsection{Breaking of Symmetry via Heterogeneity}\label{sec:break}
This symmetry need not be preserved if the two qubits are not constructed by identical projectors, as is the case in the heterogeneous architecture proposed in this paper. If projectors are non-identical, i.e., $P_\text{A} \neq P_\text{B}$, the total projection operator $\mathcal{P} = P_\text{A} \otimes P_\text{B}$ is no longer guaranteed to be symmetric:
$$ S_\text{AB} \mathcal{P} S_\text{AB}^{-1} = S_\text{AB} (P_\text{A} \otimes P_\text{B}) S_\text{AB}^{-1} = P_\text{B} \otimes P_\text{A} \neq \mathcal{P} $$
Thus, the commutator $[\mathcal{P}, S_\text{AB}]$ does not necessarily vanish and the central step in the above proof fails. The effective Hamiltonian $H'_{\text{int}} = \mathcal{P} H_{\text{int}}  \mathcal{P}$ \textit{can} become asymmetric and facilitate asymmetric operations. In fact,  we show in Sec.~\ref{sec:CNOT} that this insight can be used to implement a single-pulse fault-tolerant CNOT gate using only the natural symmetric Heisenberg exchange interaction between qubits.

To understand this in the context of QDs, below we connect the physical Fermi-Hubbard Hamiltonian provided by this quantum hardware and the effective spin Hamiltonian used in quantum information considerations.

\section{\label{sec:back}Background: The Case for Quantum Dots}

 Electron spins in semiconductor quantum dots constitute one of the leading quantum computing platforms. They offer a strong potential for scalability and long coherence times~\cite{veldhorstSiliconCMOSArchitecture2017, vandersypenInterfacingSpinQubits2017, chatterjeeSemiconductorQubitsPractice2021} and benefit from their compatibility with advanced semiconductor manufacturing techniques~\cite{steinackerIndustrycompatibleSiliconSpinqubit2025}. 
 
 This platform enables many different qubit encodings. Here we focus on the variants of singlet-triplet double quantum dot (DQD) qubits~\cite{shulmanDemonstrationEntanglementElectrostatically2012,wongHighfidelitySinglettripletST2015}, where the qubit is encoded using two neighboring quantum dots. 
 Of course, the spin of a single electron localized on one quantum dot is intrinsically a two-level system and can naturally define a (Loss-DiVincenzo) qubit~\cite{lossQuantumComputationQuantum1998}. However, controlling such a qubit can be experimentally demanding, requiring precise control of magnetic fields on the nanoscale to address individual spins~\cite{lossQuantumComputationQuantum1998}. In contrast, the mapping of a qubit onto electrons localized in two different dots allows for manipulation using electrical control, simplifying the experimental setup, and broadening the accessible operations -- at the cost of halving the number of qubits obtained from a quantum dot array~\cite{burkardSemiconductorSpinQubits2023}. Other possible encodings include exchange-only qubits that use three or four~\cite{lairdCoherentSpinManipulation2010,russQuadrupolarExchangeonlySpin2018} dots per qubit.

\subsection{From Fermi-Hubbard Model to Spin Dynamics\label{sec:Fermi-Hubbard}}
Electron spins tightly confined in gate-defined quantum dots can be modeled with the Fermi-Hubbard model, with each dot corresponding to a site that can host up to two fermions with opposite spin (due to Pauli exclusion). The Hamiltonian written in second quantization is~\cite{burkardSemiconductorSpinQubits2023}
\begin{equation}\label{eq:HFH}
    \hat{H}_\text{FH} = \sum_{i=1}^{N} \left[ \epsilon_i \hat{n}_i + U_i\hat{n}_i (\hat{n}_i - 1)\right] 
    + \sum_{\substack{\sigma=\uparrow,\downarrow \\  \braket{i,\,j}}} [ t_{ij} \hat{c}_{i\sigma}^\dagger \hat{c}_{j\sigma} + \text{H.c.} ].
\end{equation}
Here, index $i$ runs over the dots, $\epsilon_i$ is the voltage-controlled chemical potential, $\hat{c}_{i\sigma}^{\dagger}$ ($\hat{c}_{i\sigma}$) is the creation (annihilation) operator for the fermion [electron or hole] with spin $\sigma=\uparrow\text{ or }\downarrow$ defined along the spin quantization ($z$) axis. The quantity $\hat{n}_i = \sum_{\sigma=\uparrow,\downarrow} \hat{c}_{i\sigma}^\dagger \hat{c}_{i\sigma}$ is the fermionic number operator, $t_{ij}$ is the spin-conserving tunnel coupling between nearest neighbor dots, $U$ is the intradot (onsite) Coulomb repulsion, and $\text{H.c.}$ stands for Hermitian conjugate. This treatment includes the more general case of odd number of electrons per dot~\cite{barnesScreeningChargedImpurities2011} 
via renormalization of Hamiltonian parameters.  
Here,  we have omitted spin-flipping tunnel couplings, which is valid when spin-orbit coupling is weak (e.g., for electrons in Si-based quantum dots). Holes in Si or electrons in heavier materials such as $\ce{Ge}$ and $\ce{GaAs}$ can have significant spin-flip tunnel couplings, which need to be added to the second summation. Third, we have neglected any magnetic field contributions to the Hamiltonian at this stage. These will be added later. 

Experiments typically operate at the half-filled regime where the number of electrons $N$ and dots coincide. 
A single DQD qubit ($N=2$) yields 6 electronic states: $\{\ket{\uparrow,\uparrow},\ket{\uparrow,\downarrow},\ket{\downarrow,\uparrow},\ket{\downarrow,\downarrow},\ket{\uparrow\downarrow,0},\ket{0,\uparrow\downarrow}\}$ and 70 states for $N=4$, growing combinatorially fast with $\binom{2N}{N}$ states for $N$ dots. 

The Fermi-Hubbard Hamiltonian~\eqref{eq:HFH} can be mapped to the spin Hamiltonian used for quantum information~\cite{burkardSemiconductorSpinQubits2023}
\begin{equation}
H(t) = \frac{1}{4} \sum_{\langle i,j \rangle} J_{ij}(t) \boldsymbol{\sigma}_i \cdot \boldsymbol{\sigma}_j + \frac{1}{2} \sum_i g_i \mu_B \mathbf{B}_i \cdot \boldsymbol{\sigma}_i.\label{eq:HHeis}
\end{equation}
 The Heisenberg exchange (first term) defines interactions between spins $i$ and $j$ and the Zeeman splitting (second term) outlines how individual spins are affected by magnetic fields. Here, $g_i$ is the g-factor of the electron $i$, and $\mu_B$ is the Bohr magneton.
 The derivation of Eq.~\eqref{eq:HHeis} from the Fermi-Hubbard model informs us about the physical nature of the effective exchange interaction between spins~\cite{burkardCoupledQuantumDots1999}:
\begin{equation}
    \label{eq:J}
    J_{ij}=\frac{2t_{ij}^2}{U_i-\epsilon_i+\epsilon_j}+\frac{2t_{ij}^2}{U_j-\epsilon_j+\epsilon_i}.
\end{equation}
 This simplifies to $J_{ij}=4t_{ij}^2/U$ when the biases and onsite Coulomb repulsions for the two dots are equal. Note that $J_{ij}>0$, because exchange lowers the energy of the singlet relative to the triplet. Local magnetic fields $\mathbf{B}_i$ or intrinsic variability in $g_i$ enable individual addressing of qubits with magnetic fields. The latter is provided by the differences in Land\'{e} $\mathcal{G}$-tensors due to spin-orbit coupling and material inhomogeneity.

To arrive at Eq.~\eqref{eq:HHeis} from \eqref{eq:HFH}, one first assumes small bias conditions $\epsilon_i \ll U$. That is, keeping the chemical potential differences between dots much smaller than the onsite repulsion, such that each dot's occupation is unity up to small perturbative corrections. This clearly delineates the single occupation subspace of size $2^{N}$ containing all and only states with 1 fermion per site. 
The effective single occupation Hamiltonian [Eq.~\ref{eq:HHeis}] is obtained by block diagonalization via Schrieffer-Wolff perturbation theory~\cite{burkardCoupledQuantumDots1999} or, equivalently, by diagonalizing the sector formed by all singlets with a subsequent projection~\cite{geyerAnisotropicExchangeInteraction2024}. The resulting Hamiltonian describes the dynamics of spins decoupled from any dynamics of charges and has the form of a homogeneous Heisenberg interaction between spins.

\subsection{Qubit Encodings}

The single occupation subspace of the double quantum dot (DQD), $N=2$, hosts $2^N=4$ spin states, a singlet $S$ and a triplet $T_-$, $T_0$ and $T_+$.
In the presence of a strong magnetic field in the $z$-direction, the degeneracy of the triplet can be removed via Zeeman splitting. The singlet $S$ and the unpolarized triplet $T_0$ are in turn split by tunnel coupling between the two dots in the DQD.

Traditionally, the singlet and the unpolarized triplet states are chosen as the qubit states, $\ket{0}$ and $\ket{1}$, defining a computational subspace in the Hilbert space and forming the $ST_0$ qubit. 

Alternative singlet-triplet resonant ($\uparrow\downarrow$) qubits can be defined by spin of the individual electrons, i.e., $\{\ket{\uparrow\downarrow} \text{ and }\ket{\downarrow\uparrow}\}$, rather than the total spin states. Either encoding makes the qubit states resistant to global magnetic field fluctuations, since the spacing between the singlet and unpolarized triplet do not depend on the global magnetic field strength. This is the well-known decoherence-free subspace (DFS)~\cite{lidarDecoherencefreeSubspacesQuantum1998,levyUniversalQuantumComputation2002}. More recently, the use of the lower-energy triplet state ($T_-$ or $T_+$ depending on the sign of $g$) was proposed to construct~\cite{wongHighfidelitySinglettripletST2015} and experimentally realize~\cite{jirovecSinglettripletHoleSpin2021,zhangUniversalControlFour2025} a useful qubit, which we refer to as $ST_-$ below.  In all cases, the remaining levels define a leakage subspace in Hilbert space that needs to be isolated from the qubit.

\subsection{Two-qubit Gates}\label{sec:coup}

While DQD devices feature electrically controlled single-qubit operations with gate fidelities of 99.6\% in isotopically purified silicon (Si)~\cite{takedaResonantlyDrivenSinglettriplet2020}, achieving high fidelity two-qubit gates remains a significant challenge.
This type of gate is realized between qubits hosted in two neighboring DQDs that are interacting via either capacitive (Coulomb) or Heisenberg exchange coupling. The Heisenberg exchange coupling Eq.\eqref{eq:J} is controlled by manipulating the tunnel coupling $t_{ij}$ between consecutive quantum dots. The capacitive coupling is activated by controlling the charge distribution of the singlet  state by manipulating the $\epsilon_i$ through bias voltages. Motivated by a plethora of theoretical proposals~\cite{nielsenConfigurationInteractionCalculations2012,wardropExchangebasedTwoqubitGate2014,shulmanDemonstrationEntanglementElectrostatically2012,chanChargeNoiseSuppression2021}, the first exchange-based 2-qubit entangling operation was recently reported~\cite{zhangUniversalControlFour2025}. The implementation featured a variant of the $\sqrt{\text{SWAP}}$, achieving Bell state fidelities of 73-90\%. By contrast, no experimental demonstration of the CNOT gate has been reported so far in these devices, even though there is widespread interest in the quantum information community.

We argue that the disparity in progress in state-of-the-art experimental implementations of $\sqrt{\text{SWAP}}$ vs. CNOT entangling gates arise because of the symmetry of the physical setup and the underlying interqubit interaction.
Both the capacitive coupling and Heisenberg exchange arise due to electron-electron interactions, which are invariant with respect to particle permutations.  As discussed in Sec.~\ref{sec:the}, when interacting qubits are constructed using the same blueprint, this interaction symmetry leads to an invariance in the effective interaction Hamiltonian with respect to qubit exchange. In turn, this symmetry aids the implementation of  two-qubit gates that act on both qubits symmetrically,
 such as the variants of SWAP~\cite{zhangUniversalControlFour2025}, as well as XX, ZZ, and CPHASE gates~\cite{nielsenConfigurationInteractionCalculations2012,wardropExchangebasedTwoqubitGate2014,shulmanDemonstrationEntanglementElectrostatically2012,chanChargeNoiseSuppression2021} -- all requiring a simple series of pulses.
By contrast, asymmetric gates, such as CNOT or CROT,  are more difficult to achieve with capacitive coupling or with Heisenberg exchange. They 
require multiple two-qubit interactions intertwined with single-qubit gates to inject the asymmetry between the control and target qubits~\cite{levyUniversalQuantumComputation2002}, see, e.g., Refs.~\onlinecite{liControllableExchangeCoupling2012,klinovajaExchangebasedCNOTGates2012,zajacResonantlyDrivenCNOT2018,veldhorstTwoqubitLogicGate2015}. 
As we demonstrate below, high fidelity entangling CNOT gates can be generated by adopting an asymmetric device blueprint.

\section{Symmetry Constraints in Double Quantum Dot Qubits}\label{sec:model}

We now proceed to apply the principles developed in Secs.~\ref{sec:the}-\ref{sec:back} to design more efficient asymmetric quantum gates in the platform of our choice: gate-defined double quantum dot qubits. In Sec.~\ref{sec:the-1q} we establish the  notation and  specify the three qubit encodings using DQDs. Then, in Secs.~\ref{sec:the-2q} and~\ref{sec:the-2qa}  we present all six possible two-qubit interactions based on pairing these encodings.

\subsection{Single Qubit Hamiltonians\label{sec:the-1q}}

In the $\{S,T_0\}$ basis, the $ST_0$ qubit  is described via Hamiltonian~\cite{burkardSemiconductorSpinQubits2023}:
\begin{equation}
    \label{eq:H1q0}
    H_{ST_0}=\begin{bmatrix}
        -J_{12} & \Delta \zeta_{12}^{(z)}\\
        \Delta \zeta_{12}^{(z)} & 0
    \end{bmatrix},
\end{equation}
where we set the energy of $T_0$ as 0 throughout. Here we have adopted the view that the main component of the magnetic field is along the $z$-direction and defines the quantization axis. Throughout we use numbers $1,2,3,4$ etc. to label the dots.
The $\sigma_z$ control of this qubit is achieved by varying the exchange coupling  $J_{12}$ through the chemical potential of the dots $\epsilon_i$. The $\Delta \zeta_{12}^{(z)}$ generating $\sigma_x$ control is opened by magnetic field gradients created through local micromagnets and/or by inhomogeneities in the spin-orbit coupling encountered in the material in the presence of a homogeneous magnetic field.  To be general, we describe these magnetic field terms via $\boldsymbol{\zeta}_i=\frac{\mu_B}{2} \mathcal{G}_i\mathbf{B}_i$. This is a cartesian vector containing the Zeeman energy components for site $i$ along $x$, $y$ and $z$ directions, and Land\'e g-tensor $\mathcal{G}_i$ generalizing the $g_i$ of Eq.~\eqref{eq:HHeis} to magnetically inhomogeneous materials. The difference $\Delta\boldsymbol{\zeta}_{ij}=\frac{\mu_B}{2}\left(\mathcal{G}_j\mathbf{B}_j-\mathcal{G}_i\mathbf{B}_i\right)$ is between the dots $i$ and $j$.

For the resonant qubit, the Hamiltonian written in the $\{\ket{\uparrow,\downarrow},\ket{\downarrow,\uparrow}\}$ basis is
\begin{equation}
    \label{eq:Hres}
    H_\text{res}=\begin{bmatrix}
         \Delta \zeta_{12}^{(z)} -J_{12}/2 & J_{12}/2\\
         J_{12}/2 & -\Delta \zeta_{12}^{(z)} -J_{12}/2
    \end{bmatrix}.
\end{equation}
Here, the control parameters are the same as in the $ST_0$ qubit but with their roles switched. 

For the $ST_-$ qubit, we analogously get
\begin{equation}
    \label{eq:H1q-}
    H_{ST_-}=\begin{bmatrix}
        -J_{12}& \left(\Delta \zeta_{12}^{(x)}+i\Delta \zeta_{12}^{(y)}\right)/\sqrt{2}\\
        \left(\Delta \zeta_{12}^{(x)}-i\Delta \zeta_{12}^{(y)}\right)/\sqrt{2} & -\zeta_{12}^{(z)}
    \end{bmatrix}
\end{equation}
in the $\{S,T_-\}$ basis. The term $\boldsymbol{\zeta}_{ij}=\frac{\mu_B}{2}\left(\mathcal{G}_i\mathbf{B}_i+\mathcal{G}_j\mathbf{B}_j\right)$ refers to the sum over the two dots $i$ and $j$. In this case, the magnetic field along $z$ also provides $\sigma_z$ control as it shifts the $T_-$ state, while $\sigma_x$ control is achieved via the magnetic field components perpendicular to $z$.  

As summarized in Table~\ref{t:1q}, single qubit control in DQD qubits can be achieved by balancing the relative strength of the effective exchange $J_{12}$ and the particular cartesian component of the generalized magnetic field gradient $\Delta \boldsymbol{\zeta}$ between dots. Undesirable leakage outside of the computational space is caused by stray transverse magnetic field components ($\zeta_{1}^{(x,y)},\,\zeta_{2}^{(x,y)}$) in all three cases. In addition, leakage of the resonant ($\uparrow\downarrow$) and $ST_-$ qubits can be caused by the longitudinal field gradient $\Delta \zeta_{12}^{(z)}$.

\begin{table}[h]
    \centering
    \begin{tabular}{l|c|c|c}
        \toprule
        \textbf{Qubit Type } & \(\sigma_z\) control & \(\sigma_x\) control & leakage\\
        \midrule
        \(ST_0\)           & \(J_{12}\) & \(\Delta \zeta_{12}^{(z)}\) & \(\zeta_{1}^{(x,y)},\,\zeta_{2}^{(x,y)}\)\\
        \(\uparrow\downarrow\) & \(\Delta \zeta_{12}^{(z)}\) & \(J_{12}\) & \(\zeta_{1}^{(x,y)},\,\zeta_{2}^{(x,y)},\,\Delta \zeta_{12}^{(z)}\)\\
        \(ST_-\)           & \(J_{12},\,\zeta_{12}^{(z)}\) & \(\Delta \zeta_{12}^{(x,y)}\) & \(\zeta_{1}^{(x,y)},\,\zeta_{2}^{(x,y)},\,\Delta \zeta_{12}^{(z)}\)\\
        \bottomrule
    \end{tabular}
    \caption{\textbf{Single-qubit control handles for different qubit encodings.} Exchange \(J_{12}\) and the gradient of the Zeeman energy splitting vector $\boldsymbol{\zeta}_{12}$ determine the matrix elements of the effective Hamiltonian.}
    \label{t:1q}
\end{table}

\subsection{\label{sec:the-2q}Interaction Hamiltonians between Identical Qubits}

To scale the quantum hardware from a single qubit to many qubits, the standard procedure is to employ identically constructed qubits. In our context, each qubit will be constructed using a DQD in any of the three encodings discussed above. Table~\ref{t:2q} summarizes the three types of qubit coupling for each encoding in this homogeneous approach.

Coupling two $ST_0$ qubits via exchange leads to a transverse-field Ising ($-\frac{1}{4}J_{23}\sigma_{12}^x\otimes\sigma_{34}^x$) interaction between them, which can be modulated by adjusting the exchange between neighboring dots that belong to different qubits~\cite{liControllableExchangeCoupling2012}. 
Note that we continue to use numbers to refer to dots. Thus, the first qubit is formed by dots labeled by $i=1,2$ and the second qubit is formed by dots labeled by $i=3,4$. For the resonant qubits, the interqubit interaction is longitudinal-field Ising ($-\frac{1}{4}J_{23}\sigma_{12}^z\otimes\sigma_{34}^z$)~\cite{liControllableExchangeCoupling2012,wardropExchangebasedTwoqubitGate2014}, which is easy to see by changing the computational basis on both qubits. Finally, two $ST_-$ qubits are coupled via $-\frac{1}{8}J_{23}\left[\sigma^x_{12}\otimes\sigma^x_{34} + \sigma^y_{12}\otimes\sigma^y_{34} + \frac{1}{2}(1 - \sigma^z_{12})\otimes(1 - \sigma^z_{34})\right]$~\cite{zhangUniversalControlFour2025}. 

When scaling, an important consideration is that of the influence of leakage states. In $ST_0$ and resonant encodings,  the leakage states become degenerate with the computational space.  Such degeneracies can be broken by introducing local magnetic fields, which is experimentally challenging. By contrast, the $ST_-$ encoding isolates the computational space energetically, making this encoding preferable for multi-qubit devices, e.g. Refs.~\onlinecite{zhangUniversalControlFour2025,geyerAnisotropicExchangeInteraction2024}.

\begin{table}[h]
    \centering
    \renewcommand{\arraystretch}{1.8} 
    \begin{tabular}{l c l}
        \toprule
        \thead{\textbf{Coupled}\\\textbf{Qubits}} & \textbf{Coupling} & \thead{\textbf{Degeneracies with}\\\textbf{leakage states}} \\
        \midrule
        
        $ST_0 + ST_0$ & $-\frac{1}{4}J_{23}\sigma^x_{12} \otimes \sigma^x_{34}$ & 
        \makecell[l]{$\ket{T_0,T_0} \rightleftharpoons$ \\ $\ket{T_-,T_+}, \ket{T_+,T_-}$} \\
        
        $\uparrow\downarrow + \uparrow\downarrow$ & $-\frac{1}{4}J_{23}\sigma^z_{12} \otimes \sigma^z_{34}$ & 
        \makecell[l]{$\ket{\uparrow\downarrow,\uparrow\downarrow} \text{, etc.} \rightleftharpoons$ \\ $\ket{\uparrow\uparrow,\downarrow\downarrow}, \ket{\downarrow\downarrow,\uparrow\uparrow}$} \\
        
        $ST_- + ST_-$ & 
        $\begin{aligned} -\frac{1}{8}J_{23} \big[ &\sigma^x_{12}\otimes\sigma^x_{34} + \sigma^y_{12}\otimes\sigma^y_{34} \\
                                       & + \frac{1}{2}(1 - \sigma^z_{12})\otimes(1 - \sigma^z_{34})\big] \end{aligned}$ & 
        None \\
        
        \bottomrule
    \end{tabular}
    \caption{\textbf{Coupling qubits with the same encoding leads to \textit{symmetric} interactions.} Qubit A (B) is formed by dots 1 and 2 (3 and 4). In $ST_0$ and resonant qubits an interqubit magnetic field gradient $\Delta \zeta^{(z)}_{23}$ is needed to prevent leakage due to degeneracies between computational (e.g. $\ket{T_0,T_0}$) and leakage states (e.g. $\ket{T_-,T_+}$.)}
    \label{t:2q}
\end{table}

Overall, we emphasize that the interaction between identically constructed qubits is symmetric with respect to qubit exchange. 
In contrast, the CNOT gate--the basic building block of many quantum algorithms--is manifestly non-symmetric. As a result, additional interactions are needed to break the exchange symmetry resulting in more complex schemes to construct asymmetric gates~\cite{liControllableExchangeCoupling2012,klinovajaExchangebasedCNOTGates2012}.

\subsection{Our Proposal: The Case for Non-identical Qubit Encodings}\label{sec:the-2qa}

Previous work has attempted to utilize the individual strengths of distinct qubit platforms in heterogeneous qubit encodings, while also recognizing the challenges~\cite{mehlSimpleOperationSequences2015}. For instance, Ref.~\onlinecite{noiriFastQuantumInterface2018}  demonstrates the coupling interface that is useful for a CPHASE gate by pairing a Loss-DiVincenzo and a singlet-triplet qubit in which the faster qubit initialization and readout for singlet-triplet qubits is exploited. Here we go beyond these initial arguments, by investigating the emerging properties of heterogeneous qubit arrays that arise from broken symmetries. That is, by investigating how non-identical qubit encodings are different from the sum of their parts.

Table~\ref{t:2qNew} summarizes three pairings of distinctly encoded qubits. These were obtained via Schrieffer-Wolff perturbation theory~\cite{burkardCoupledQuantumDots1999} for the system of four dots analogously to the results from the literature presented in Sec.~\ref{sec:the-2q}. 
As opposed to those in Table~\ref{t:2q}, all of them result in a distinct asymmetric interqubit interaction. 
The most straightforward way to break the qubit-exchange symmetry 
is to rotate the basis of one of the qubits relative to the uniform resonant or $ST_0$ mapping. This 1-qubit basis change results in an $ST_0+\uparrow\downarrow$ pair with $XZ$ interaction that can facilitate the implementation of the asymmetric CNOT gate. However, this mapping may present experimental challenges, as $ST_0$ vs. resonant qubits require different control pulses. The $ST_0$ is manipulated using baseband voltage pulses applied to the gates, while the resonant qubit by excitation using external microwave sources.

By contrast, by pairing the $ST_0$ and $ST_-$ qubits an asymmetric $XZ$ interaction between qubits is achieved 
without introducing the necessity for the microwave radiation, simplifying the experimental set-up as both singlet-triplet qubits are controlled by electrostatic gates. In fact, as we detail below, the $ST_-$ and $ST_0$ pairing of qubits leads to an experimentally realistic platform for the realization of a fast single-pulse fault-tolerant CNOT gate.

Lastly, the pairing of $ST_-$ with the resonant qubit yields an interaction that affects the phases of the two qubits differently, suggesting that it could potentially be useful for controlled rotation (CROT) gates~\cite{geyerAnisotropicExchangeInteraction2024}. Moreover, in this case the computational basis states are no longer degenerate with any of the leakage states, which could potentially alleviate leakage relative to the resonant qubit devices. Nevertheless, this type of mixed encoding has the same experimental challenges as the $ST_0+\uparrow\downarrow$, and does not merit further consideration at this stage.

\begin{table}[htb]
    \centering
\begin{tabular}{l|c|c}
    \toprule
    \textbf{Coupled } & coupling & degeneracies\\\textbf{Qubits}&&with leakage states\\
    \midrule
    \(ST_-+ST_0\) & \(-\frac{1}{8}J_{23}(1-\sigma^z_{12})\otimes\sigma^x_{34}\) & \(\ket{T_-,T_0}\rightleftharpoons\ket{T_0,T_-},\text{etc.}\)\\
    \(ST_0+\uparrow\downarrow\) & \(-\frac{1}{4}J_{23}\sigma^x_{12}\otimes\sigma^z_{34}\) & \(\ket{T_0,\uparrow\downarrow}\rightleftharpoons\ket{T_-,\uparrow\uparrow},\ket{T_+,\downarrow\downarrow}\)\\
    \(ST_-+\uparrow\downarrow\) & \(-\frac{1}{8}J_{23}(1-\sigma^z_{12})\otimes\sigma^z_{34}\) & None\\
        \bottomrule
\end{tabular}
    \caption{\textbf{Coupling qubits with different encodings leads to \textit{asymmetric} interactions.} Pairing of either $ST_-$ or a resonant qubit with an $ST_0$ qubit leads to an $XZ$ interaction, that facilitates the realization of CNOT. While degeneracies are introduced as in Table~\ref{t:2q}, leakage suppression strategy is unchanged. Pairing an $ST_-$ and a resonant qubit gives an asymmetric phase shift. 
    }
    \label{t:2qNew}
\end{table}

\section{\label{sec:CNOT}Single-pulse CNOT Gate via the $ST_-+ST_0$ Pairing}

As established in Sec.~\ref{sec:the-2qa} (Table~\ref{t:2qNew}), pairing an $ST_-$ with an $ST_0$ qubit yields an asymmetric $XZ$-type interqubit coupling, $-\frac{1}{8}J_{23}(1-\sigma^z_{12})\otimes\sigma^x_{34}$, while retaining the experimental simplicity of electrostatic control. We now exploit this asymmetry to construct a direct, single-pulse CNOT gate and benchmark its performance under realistic experimental conditions. 

In Sec.~\ref{sec:the-cnot} we derive the effective two-qubit Hamiltonian and in Sec.~\ref{sec:ideal} we derive the timing conditions 
needed to realize the CNOT gate and demonstrate its ideal action on Bell states. Finally, we test the gate against two dominant sources of infidelity in gate-defined quantum dots: leakage out of the computational subspace (Sec.~\ref{sec:leakage}) and quasi-static electric and magnetic noise (Sec.~\ref{sec:Noise}).

\subsection{\label{sec:the-cnot}Effective Hamiltonian and Gate Construction}

The upper-triangular portion of the two-qubit Hamiltonian for an $ST_-+ST_0$ device, written in the basis $\{\ket{S,S},\ket{S,T_0},\ket{T_-,S},\ket{T_-,T_0}\}$, reads
\begin{widetext}
\begin{equation}
\label{eq:H-0}
    H^{ST_-+ST_0}=H_{12}^{ST_-} + H_{34}^{ST_0} - \frac{1}{8}J_{23}(1-\sigma^z_{12})\otimes\sigma^x_{34}=\begin{bmatrix}
    \zeta^{(z)}_{12} -J_{12}-J_{34} & \Delta\zeta^{(z)}_{34} & \left(\Delta \zeta^{(x)}_{12}+i\Delta \zeta^{(y)}_{12}\right)/\sqrt{2} & 0\\
    * &\zeta^{(z)}_{12} - J_{12} & 0 & \left(\Delta \zeta^{(x)}_{12}+i\Delta \zeta^{(y)}_{12}\right)/\sqrt{2}\\
    * & * & -J_{34} & \Delta\zeta^{(z)}_{34} - J_{23}/4\\
    * & * & * & 0
    \end{bmatrix}.
\end{equation}
\end{widetext}
To derive Eq.~\eqref{eq:H-0} we have followed the Schrieffer-Wolff transformation~\cite{burkardCoupledQuantumDots1999} for defining low-energy subspace Hamiltonian, which naturally leads to the Hamiltonians of the individual qubits defined in \eqref{eq:H1q0} and \eqref{eq:H1q-} and their interaction. 
In this qubit-pair, individual qubit control is provided by the intraqubit exchanges $J_{12}$ and $J_{34}$, magnetic field in the $z$-direction and the effective magnetic-field gradients $\Delta\boldsymbol{\zeta}_{ij}$. 
The distinguishing feature of Eq.~\eqref{eq:H-0} relative to the cases of symmetrically interacting qubits (Table~\ref{t:2q}) is that the interqubit exchange $J_{23}$ appears only once in the upper-triangular block, as an off-diagonal element coupling $\ket{T_-,S}$ and $\ket{T_-,T_0}$ but not $\ket{S,S}$ and $\ket{S,T_0}$. This singular appearance is a direct consequence of the asymmetric $XZ$ coupling and --- as we show next --- is what enables a direct single-pulse implementation of CNOT.

For the duration of the gate, the intraqubit magnetic-field gradients are turned off, $\Delta\boldsymbol{\zeta}=\mathbf{0}$. 
For convenience we will now set the zero of energy to the state $\ket{T_-,T_0}$, reducing the Hamiltonian~\eqref{eq:H-0} to
\begin{equation}
\label{eq:HCNOT}
    H_\text{2q}=\begin{bmatrix}
        T_\text{A} + T_\text{B} & 0 & 0 & 0\\
        0 & T_\text{A} & 0 & 0\\
        0 & 0 & T_\text{B} & T_\text{AB}\\
        0 & 0 & T_\text{AB} & 0
    \end{bmatrix},
\end{equation}
where $T_\text{A}=\zeta_{12}^{(z)}-J_{12}$ is the level splitting of the $ST_-$ qubit (qubit A), $T_\text{B}=-J_{34}$ is the level splitting of the $ST_0$ qubit (qubit B), and $T_\text{AB}=-J_{23}/4$ is the interqubit coupling. 

It is now convenient to transition from physical  to logical qubit labels.  The singlet (triplet) state is assigned the label $\ket{0}$       ($\ket{1}$) such that, for example, $\ket{T_-,T_0}= \ket{1_\text{A},1_\text{B}}$. The basis ordering in Eq. \eqref{eq:H-0} is identical to the one in \eqref{eq:HCNOT} but is now labeled as  $\ket{0_\text{A},0_\text{B}}$, $\ket{0_\text{A},1_\text{B}}$, $\ket{1_\text{A},0_\text{B}}$, $\ket{1_\text{A},1_\text{B}}$.

The structure of $H_\text{2q}$ exposes the conditional logic required for CNOT. When A is in $\ket{0_\text{A}}$, qubit B sees only the diagonal elements and does not rotate; when A is in $\ket{1_\text{A}}$, the off-diagonal $T_\text{AB}$ couples $\ket{0_\text{B}}$ and $\ket{1_\text{B}}$ and drives a bit flip. This is precisely the action of a control--target gate: 
the CNOT gate flips the target qubit if and only if the control qubit is in $\ket{1}$. That is,
\begin{equation}
\text{CNOT}=e^{i\frac{\pi}{4}(I_1-Z_1)(I_2-X_2)}=\begin{bmatrix}
1 & 0 & 0 & 0\\
0 & 1 & 0 & 0\\
0 & 0 & 0 & 1\\
0 & 0 & 1 & 0
\end{bmatrix}.
\label{eq:cnot}
\end{equation}

\subsection{Ideal Gate Operation}\label{sec:ideal}

We first consider the limiting case in which $|T_\text{B}|$ is small relative to $|T_\text{AB}|$ (i.e.\ $T_\text{B}\rightarrow0$), where the fidelity of the CNOT gate is maximized; the effect of a residual finite $T_\text{B}$ is examined in Sec.~\ref{sec:tabnonz}. In this limit, comparing the reduced Hamiltonian~\eqref{eq:HCNOT} with the target gate~\eqref{eq:cnot} shows that CNOT can be generated by a single pulse of
\begin{equation}
\label{eq:HCNOTfinal}
H_\text{CNOT}=\begin{bmatrix}
T_\text{A} & 0 & 0 & 0\\
0 & T_\text{A} & 0 & 0\\
0 & 0 & 0 & T_\text{AB}\\
0 & 0 & T_\text{AB} & 0
\end{bmatrix},
\end{equation}
in which qubit B idles unless driven by qubit A. By inspection, the propagator $U(t)=e^{-iH_\text{CNOT}t}$ is
\begin{equation}
\label{eq:UCNOT}
U(t)=\begin{bmatrix}
e^{-iT_\text{A} t} & 0 & 0 & 0\\
0 & e^{-iT_\text{A} t} & 0 & 0\\
0 & 0 & \cos(T_\text{AB} t) & -i\sin(T_\text{AB}t)\\
0 & 0 & -i\sin(T_\text{AB}t) & \cos(T_\text{AB}t)
\end{bmatrix}.
\end{equation}
When the control is in $\ket{0}_\text{A}$ the target merely accumulates the phase $e^{-iT_\text{A}t}$; when the control is in $\ket{1}_\text{A}$ the target rotates about $\hat{x}$ at rate $T_\text{AB}$.

To recover the CNOT of Eq.~\eqref{eq:cnot} up to an irrelevant global phase, we require the following two conditions for the gate time $\tau$:
\begin{enumerate}
    \item $\tau\times T_\text{AB}=\frac{\pi}{2}+\pi k_\text{AB}$,
    \item $\tau\times T_\text{A} = \pi(2 k_\text{A}-k_\text{AB})+\frac{\pi}{2}$,
\end{enumerate}
with $k_\text{A},\,k_\text{AB}\in\mathbb{Z}$ (integers). Condition~1 makes the conditional rotation an odd multiple of $\pi/2$, ensuring a full bit flip on the target with $\cos(T_\text{AB}t)=0$; condition~2 returns the control qubit to itself with the matching global phase $\pm i$. 
\begin{figure}[b]
\includegraphics[width=0.49\textwidth]{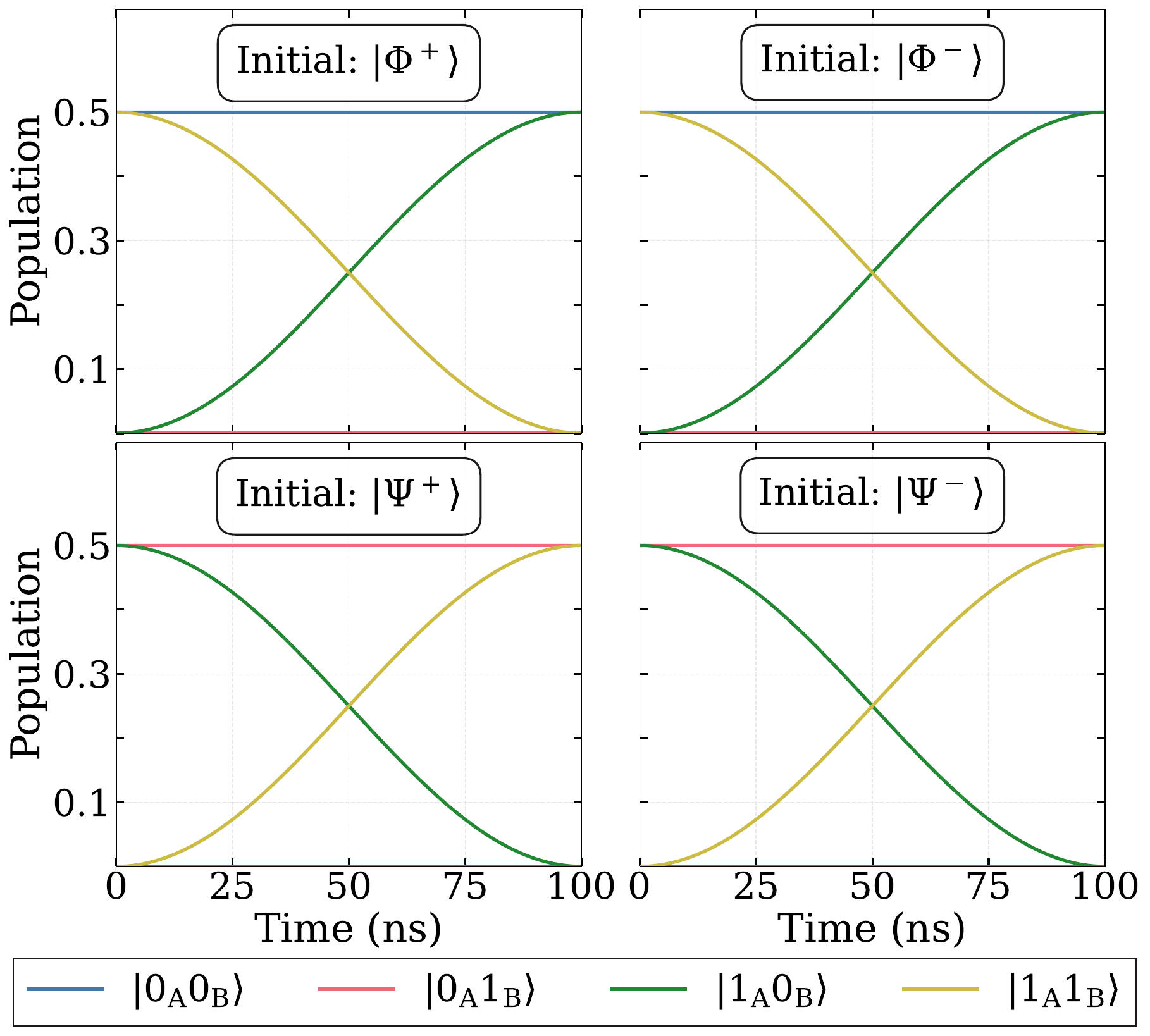}%
\caption{\label{fig:bell} \textbf{Ideal CNOT gate operation in a heterogeneous qubit encoding.} Four Bell states ($\ket{\Phi^\pm} = \frac{1}{\sqrt{2}}(\ket{0_\text{A}0_\text{B}} \pm \ket{1_\text{A}1_\text{B}})$, $\ket{\Psi^\pm} = \frac{1}{\sqrt{2}}(\ket{0_\text{A}1_\text{B}} \pm \ket{1_\text{A}0_\text{B}})$) are evolved under the Hamiltonian~\eqref{eq:HCNOTfinal} over $100$~ns. Each panel shows the population of each computational basis state $\{\ket{0_\text{A}0_\text{B}},\ket{0_\text{A}1_\text{B}},\ket{1_\text{A}0_\text{B}},\ket{1_\text{A}1_\text{B}}\}$. The populations of the computational states with leading $0$ are unaffected and the populations of the states with leading $1$ are interchanged.}
\end{figure}

Figure~\ref{fig:bell} demonstrates the action of Hamiltonian~\eqref{eq:HCNOTfinal} on the four Bell states. Since the evolution is unitary, we work with pure states and plot the projection onto each computational basis state. The interqubit coupling is set to $J_\text{23}=2\pi\times 10^7$~rad/s (corresponding to experimentally measured frequency of $10$~MHz)\cite{reedReducedSensitivityCharge2016,connorsChargenoiseSpectroscopySi2022}, and the timing conditions above are imposed with $k_\text{AB}=k_\text{A}=-1$ (the shortest gate time). This fixes $T_\text{AB}=T_\text{A}=-5\pi \times 10^6$~rad/s and the gate time at $\tau=100$~ns. Each Bell state transforms as expected under a perfect CNOT.

This figure was generated for a square pulse, without taking into account the finite ramp-up and ramp-down of $T_\text{AB}$. However, we have verified that both sinusoidal and linear ramping yield a perfect CNOT with timing conditions determined by the time-averaged value of $T_\text{AB}$.

Demonstrating the ideal limit, however, is only the first step. To assess whether this proposal is viable for realistic gate-defined quantum dot devices, we must account for three sources of gate infidelity:  (i) leakage outside the computational subspace,  (ii) nonzero exchange coupling in qubit B ($J_{34}=-T_\text{B}\neq 0$), and (iii) decoherence due to imperfect isolation from the environment. We address these  below.

\subsection{\label{sec:leakage}Effect of Leakage}
\begin{figure}[b]
\includegraphics[width=0.49\textwidth]{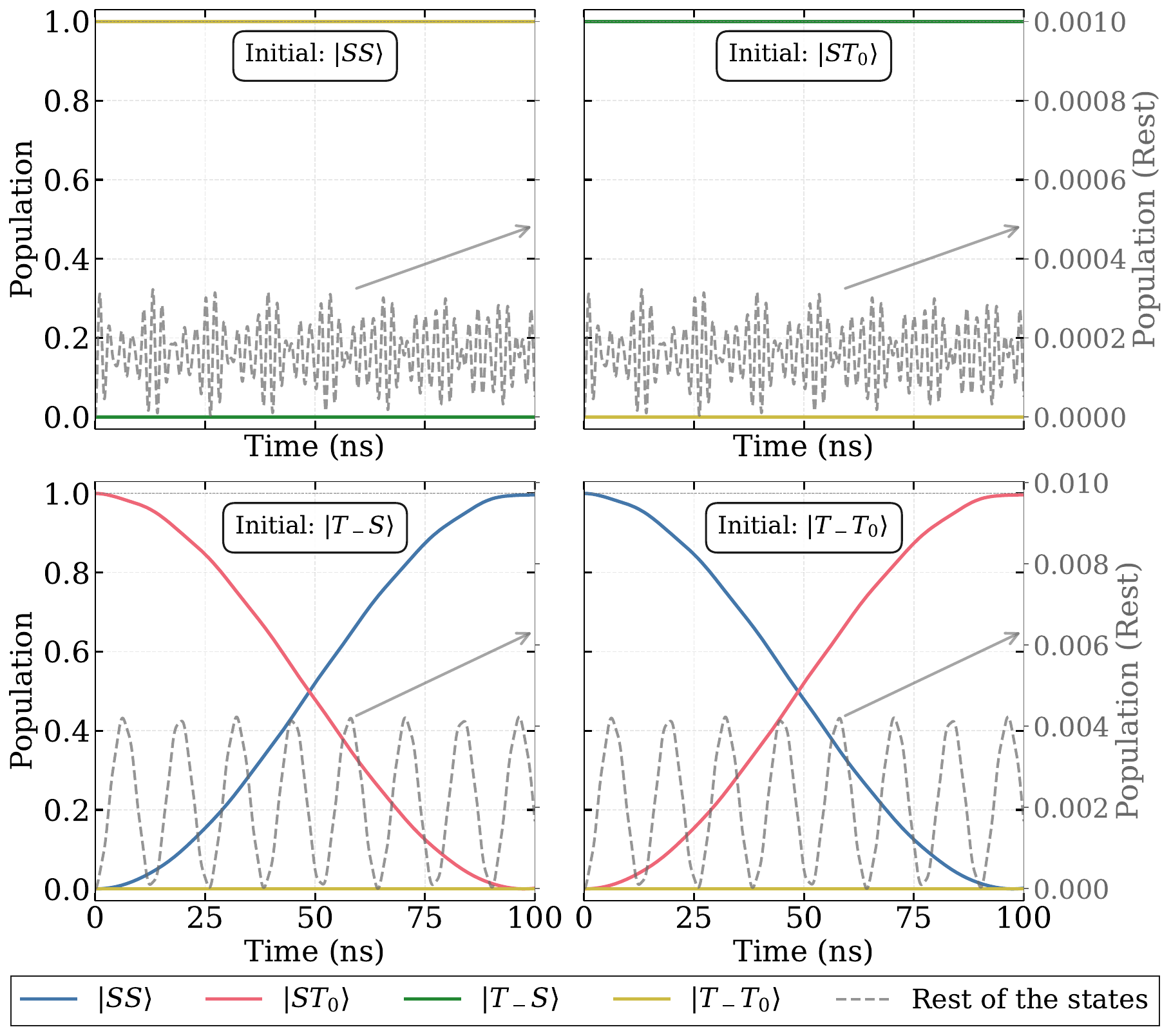}
\caption{\label{fig:leakage} \textbf{CNOT in the presence of leakage states.} Time evolution over the gate duration of $100$~ns of the four-spin system. Note that the population outside of the computational space is plotted in gray on a secondary $y$-axis which is zoomed in by a factor of 1000 (panels a,b) and a factor of 100 (panels c,d).}
\end{figure}

For a two-qubit gate, the computational space consists of 4 out of the 16 single occupation manifold levels, with 12 being leakage states. We verify that CNOT operation does not cause considerable leakage outside of the computational space in Figure~\ref{fig:leakage}. To do so, we propagate the dynamics under the full 16-level Hamiltonian as determined by Eq.~\eqref{eq:HHeis}.
Echoing the ideal example (Sec.~\ref{sec:ideal}), we keep the level splitting of the $ST_-$ qubit $T_\text{A}=\zeta^{(z)}_{12}-J_{12}=-5\pi\times10^6$ rad/s and $T_\text{B}=-J_{34}=0$, and the interqubit coupling constant at $T_\text{AB}=-J_{23}/4 = -5\pi\times10^6$~rad/s. 
The magnetic fields are oriented along the $z$-axis with $\zeta^{(z)}_{1}=\zeta^{(z)}_{2}=2\pi\times10^9$ rad/s ($1$~GHz) and $\zeta^{(z)}_3=\zeta^{(z)}_4=1.7\pi\times10^9$ rad/s ($850$~MHz), which are representative of the usual experimental conditions~\cite{wuTwoaxisControlSinglettriplet2014,zhangUniversalControlFour2025,takedaResonantlyDrivenSinglettriplet2020}.
The non-zero magnetic field in the $z$-direction is required to break the triplet degeneracy and suppress the leakage into the $T_+$ states. It is also necessary to establish a non-zero magnetic field gradient in the $z$ direction between the dots to suppress the leakage into $\ket{T_0,T_-}$.

Each panel of Fig.~\ref{fig:leakage} shows the system initialized in one of the four computational basis states and evolved over a period of $100$~ns under the influence of the CNOT gate pulse. Because the populations of leakage states are orders of magnitude smaller than that of the computational states, we utilize a secondary $y$-axis to show the sum of populations in all the leakage states in gray. When the control qubit is in the state $\ket{0}$, the two-qubit state remains unchanged, apart from the leakage of less than $4\times10^{-4}$. When the control qubit is in the state $\ket{1}$, leakage of up to $4\times10^{-3}$ accompanies the bit flip of the second qubit. In this case, the slightly higher leakage is due to transitions to $\ket{T_0T_-}$ and $\ket{ST_-}$, i.e. the leakage states that have the same structure as the computational basis states but opposite qubit ordering.

We thus conclude that the magnetic field gradient $\Delta \zeta^{(z)}_{23}$ of realistic magnitude ($150$ MHz) efficiently suppresses the leakage during this novel implementation of the CNOT gate.

\subsection{\label{sec:tabnonz}Influence of $T_\text{B}\neq0$}

\begin{figure}[b]
\includegraphics[width=0.49\textwidth]{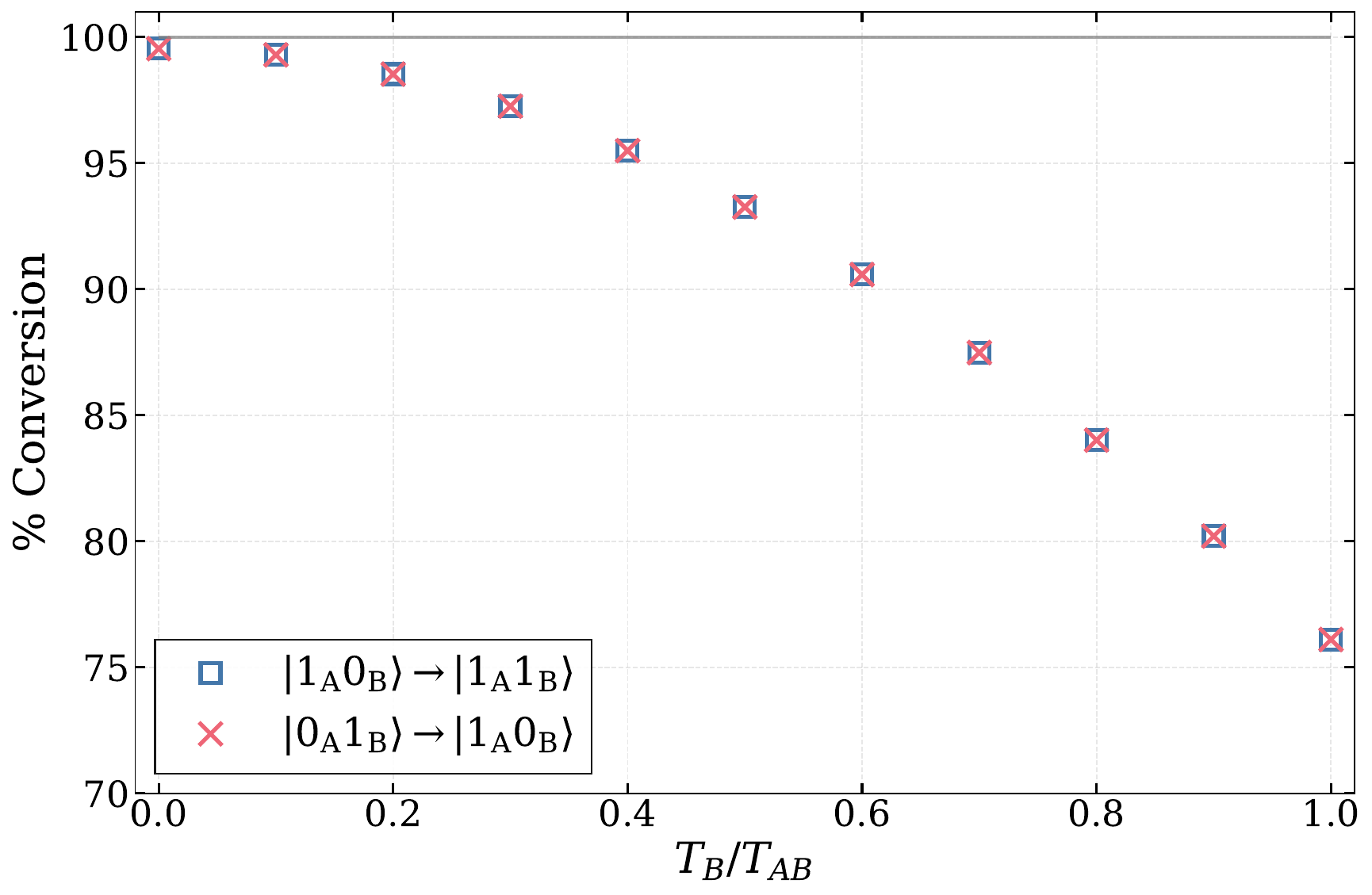}
\caption{\label{fig:TB}\textbf{Effect of non-zero $T_\text{B}$.} Percent interconversion between the $\ket{10}$ and $\ket{11}$ upon applying the CNOT gate as a function of $T_\text{B}$ relative to $T_\text{AB}$. Note the quadratic decay of the percent conversion, enabling essentially full interconversion for small values of $T_\text{B}$.}
\end{figure}

We now assess the influence of $T_\text{B}\neq 0$ on the CNOT gate. For this, we increase the $|T_\text{B}|$ magnitude while keeping the gate time $\tau$ and other gate parameters intact.  Figure~\ref{fig:TB} shows the percent of $\ket{10} \leftrightarrow \ket{11}$ interconversion with increasing $|T_\text{B}/T_\text{AB}|$. The simulations are performed with the same parameters as in Fig.~\ref{fig:leakage}, except $T_\text{B}$ is allowed to vary. The leftmost point corresponds to $T_\text{B}=0$, featuring the state interconversion of 99.6\% shown in Fig.~\ref{fig:leakage}. 
The conversion probability decays as $ \sim |T_\text{B}/T_\text{AB}|^2$, with the linear term being negligible in the range considered. For this reason, increasing $|T_\text{B}|$ to as much as 10\% of $|T_\text{AB}|$ leaves the effectiveness of the gate essentially intact, with a reduction of the state interconversion of no more than 0.2\%. 

When $|T_\text{B}|\ne 0$ it breaks the degeneracy between $\ket{0_\text{A},0_\text{B}}$ and $\ket{0_\text{A},1_\text{B}}$, and that between  $\ket{1_\text{A},0_\text{B}}$ and $\ket{1_\text{A},1_\text{B}}$. This leads to a shift of the transition frequency that transitions the originally designed gate (for $T_\text{B}=0$) off-resonance, and to mixing of the computational states. 
The detrimental influence of  $T_\text{B}\ne 0$ on gate operation can be further minimized  by correcting for the frequency shift, changing the gate timing,  and accounting for the coherent gate errors that result from the computational state mixing. Thus, the results shown here overestimate the influence of $T_\text{B}$ on gate operation, and should be seen as an upper limit.

Overall, these results show that small $|T_\text{B}/T_\text{AB}|<0.1$ are tolerated while maintaining high gate fidelity. This requirement of  small $|T_\text{B}/T_\text{AB}|$ can be achieved by controlling the gate voltages, in a manner akin to what was already accomplished in Ref.~\onlinecite{zhangUniversalControlFour2025} for nonzero magnetic field gradients that are made negligibly small relative to the intraqubit exchange coupling.

\subsection{\label{sec:Noise}Effect of Quasi-Static Noise}
We next address the effect of the environmentally-induced decoherence on the performance of the CNOT gate. In semiconductor quantum dots, it is the low-frequency components of electrical and hyperfine noise that dominate. Because this source of quantum noise is essentially frozen during the dynamics, a quasi-static treatment is appropriate~\cite{burkardSemiconductorSpinQubits2023}. In this approach, the noise is treated as a source of inhomogeneity and sampled from a Gaussian distribution to define parameters for an individual quantum realization, which are then averaged to yield the density matrix influenced by quasi-static noise.

For these simulations, we keep the average interqubit coupling $T_\text{AB}= -J_{23}/4 = -5\pi\times10^6$~rad/s but apply a sinusoidal ramp-on profile~\cite{wardropExchangebasedTwoqubitGate2014} that starts and ends with $T_\text{AB}=0$ over the $100$~ns duration of the pulse as in Fig.~\ref{fig:leakage}. Noise is introduced in both the $z$-component of the magnetic field and the exchange couplings: isotropic Gaussian magnetic-field fluctuations with mean $0$ and standard deviation $0.44\pi\times10^6$~rad/s per Cartesian component --- corresponding to a single-spin inhomogeneous dephasing time $T_2^*\approx 3.6\,\mu$s --- and Gaussian fractional exchange noise with standard deviation $(2\pi Q)^{-1}$, with an exchange-oscillation quality factor $Q=40$, which is accessible in nuclear-spin free systems like $\ce{^{28}Si}$~\cite{dialChargeNoiseSpectroscopy2013,nicholHighfidelityEntanglingGate2017,yonedaQuantumdotSpinQubit2018,mauneCoherentSinglettripletOscillations2012}. 
Leakage is also taken into account because the simulation spans both the computational and leakage states. We further consider the cases of $T_\text{B}=0$ and $T_\text{B}\neq0$.

\begin{figure}[b]
(a) \includegraphics[width=0.45\textwidth]{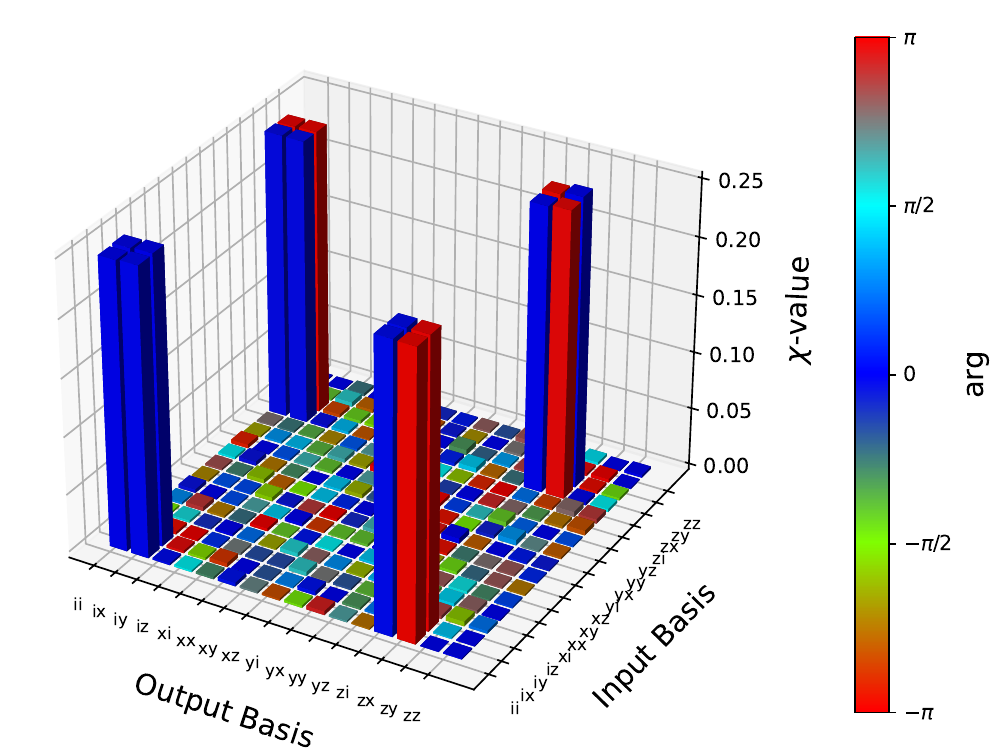}
(b) \includegraphics[width=0.45\textwidth]{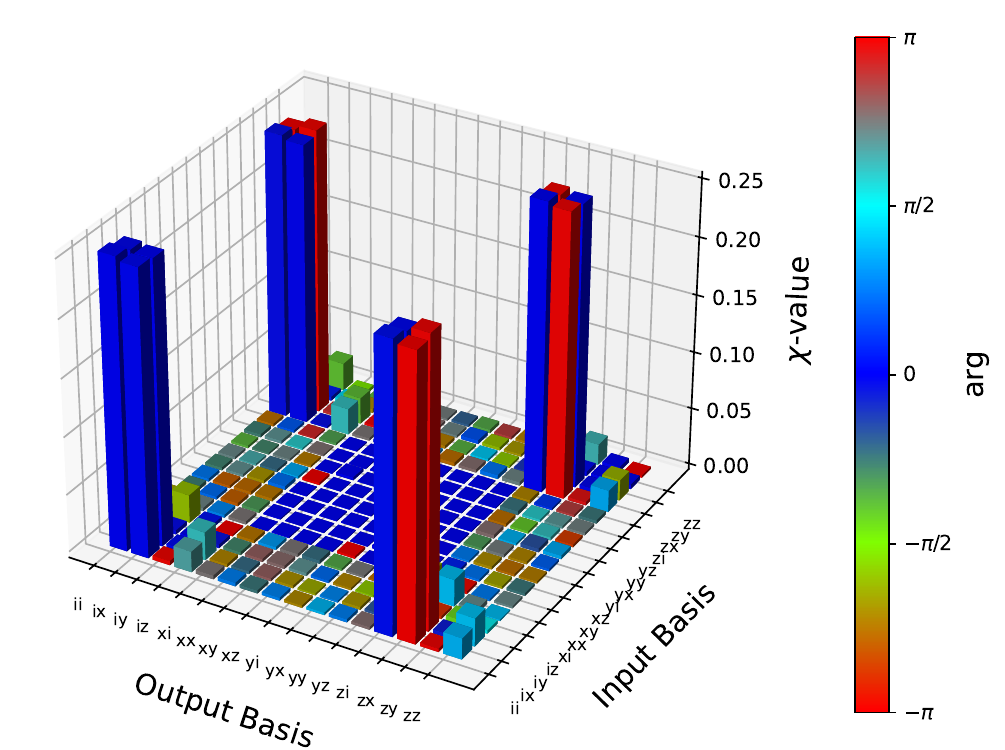}
\caption{\label{fig:Noise}  \textbf{The $\chi$-matrix characterizing the proposed CNOT pulse} applied to a pair of double quantum dots with (a) $T_\text{B}=0$ featuring $99.6\%$ gate fidelity and (b) $T_\text{B}=0.1T_\text{AB}$ with $99.2\%$ gate fidelity. The parameters in Eq.~\ref{eq:HHeis} are: $T_\text{A}=-5\pi\times10^6$~rad/s,  $\zeta^{(z)}_{12}=1.3\pi\times10^9$~rad/s and $\zeta^{(z)}_{34}=\pi\times10^9$~rad/s ($650$ and $500$ MHz), $T^\text{av}_\text{AB}=-5\pi\times10^6$~rad/s. Noise magnitudes are set by $Q=40$ for the exchange noise on each nearest-neighbor dot pair (1--2, 2--3, 3--4) and $\Delta B_i^{(z)}=0.44\pi\times10^6$~rad/s for the magnetic noise. Note that the 0.4\% decrease in fidelity for panel (b) is mostly due to the coherent error.}
\end{figure}

To characterize the two-qubit CNOT operation, Figure~\ref{fig:Noise} reconstructs the process matrix (or $\chi$-matrix) for the proposed CNOT gate for (a) $T_\text{B}= 0$ and (b) $T_\text{B}= 0.1T_\text{AB}$.
Each element of the matrix is given by how the corresponding two-qubit tensor-product Pauli basis element (e.g. $YZ= Y_\text{A} \otimes Z_\text{B}$ where $I$, $X$, $Y$ and $Z$ are the usual Pauli matrices) transforms under the action of this operation. The ideal CNOT gate has 16 non-zero elements, all equal to $\pm1/4$.  Specifically, only the Pauli components $II$, $IZ$, $XI$, and $XZ$ contribute, reflecting the decomposition $U_\text{CNOT}=(II+IZ+XI-XZ)/2$. Nonzero elements outside this ideal pattern quantify imperfections in the implemented gate that may arise due to noise, and that reduce gate fidelity.

The gate quality is characterized via process fidelity that is defined as
\begin{equation}
    \mathcal{F}=\Tr \left[\chi^\dagger_\text{ideal}\chi_\text{real}\right].
\end{equation}
The calculated fidelities are $99.6\%$ for $T_\text{B}=0$ and $99.2\%$ for $T_\text{B}=0.1T_\text{AB}$. The decrease in fidelity of $0.4\%$ for $T_\text{B}\ne 0$ is minor, keeping the gate within the fault-tolerant threshold of 99\% for two-qubit gates and is largely attributed to the incomplete rotation. The coherent error caused by $T_\text{B}\neq0$ can be further reduced as discussed in Sec.~\ref{sec:tabnonz}. Thus, the proposed hybrid $ST_-+ST_0$ platform operates the CNOT gate in the fault-tolerant regime under realistic, experimentally accessible conditions, confirming the heterogeneous-architecture strategy introduced in Sec.~\ref{sec:the-2qa} as a practical route to asymmetric two-qubit gates. 

\subsection{Effect of the Magnetic Field Strength}
\begin{figure}[b]
\includegraphics[width=0.49\textwidth]{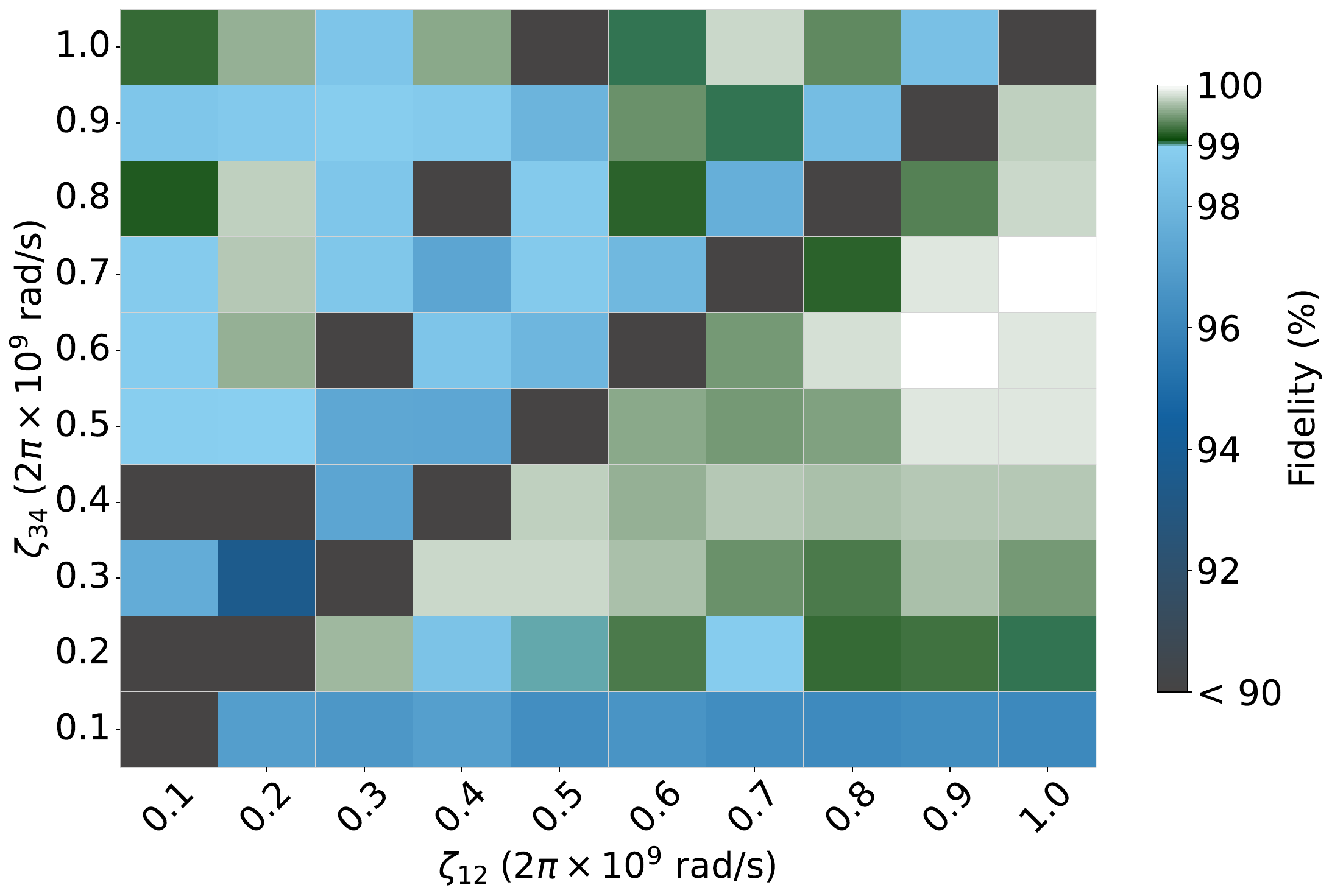}
\caption{\label{fig:B1B2}\textbf{Fidelity of the CNOT for different magnetic fields} on qubit $A$, $\zeta^{(z)}_\text{12}$ and qubit $B$, $\zeta^{(z)}_\text{34}$. The operation is robust to the particular magnetic field strength as long as the magnetic field and the gradient are both non-zero.}
\end{figure}

As discussed in Secs.~\ref{sec:the-2q} and~\ref{sec:leakage}, magnetic field gradients between qubit $A$ and $B$ are needed to suppress the leakage out of the computational states during this novel implementation of the CNOT gate. The gradient is necessary to break the degeneracy between the computational triplet states and the leakage state with the triplet ordering interchanged (e.g. $T_-T_0$ and $T_0T_-$). 

Figure~\ref{fig:B1B2} shows the fidelity of the CNOT gate as a function of magnetic field strengths on the two individual qubits (dots 1 and 2 encode the $ST_-$ qubit; 3 and 4 the $ST_0$). For simplicity, we focus on $T_\text{B}=0$, with $T_\text{B}\ne0$ results being very similar. Lighter colors signify higher fidelities with fidelities above 99\% shown in green, those between 90\% and 99\% in blue, and those below 90\% in gray. In the absence of magnetic field gradients (diagonal), gate fidelities are poor. 
By contrast, gate fidelity in the high 90's is achieved for various magnetic field strengths so long as there is a sufficiently strong magnetic field gradient between the qubits (off-diagonals).  Note that the magnetic field gradient suppresses the leakage more effectively for a particular direction of magnetic field gradient, i.e. when the $ST_-$ qubit is subject to stronger magnetic field, as reflected by the asymmetry of the plot. This is because the stronger magnetic field does not meaningfully affect the $ST_0$ qubit; in contrast, on the $ST_-$ qubit, stronger field better suppresses the leakage to the $T_+$ state. This magnetic field gradient directionality cannot make a difference in the case of homogeneous mapping as the qubits are identical. Crucially, the CNOT gate fidelity is robust for a range of magnetic field strengths off-diagonal in the lower triangular portion of the plot.

\section{\label{sec:Disc}Discussion}
\subsection{Implementation in Double Quantum Dots}
\label{sec:Params}

The importance of having a ``simple" CNOT is hard to overestimate as it is heavily featured in many quantum algorithms, and is needed to form the textbook example of the universal gate set~\cite{nielsenQuantumComputationQuantum2011}.  Successful qubit platforms often feature a ``simple" CNOT. For example, the cross-resonance gate often implemented in superconducting qubits~\cite{rigettiFullyMicrowavetunableUniversal2010} requires only one two-qubit interaction and one additional local $\pi/2$ rotation of each qubit to realize a CNOT gate.

Our proposal for the realization of a ``simple" high-fidelity CNOT gate in double quantum dots by designing an alternating $ST_-$-$ST_0$ heterogeneous $AB$ qubit architecture can be implemented straightforwardly in existing DQD devices, as the nature of each qubit can be tuned through gate voltages.  In fact, we have shown in Sec.~\ref{sec:Noise} that the gate can be operated above the fault-tolerance threshold of 99\% under realistic experimental conditions. 

Previously, there had been other proposals to employ hybrid architectures~\cite{mehlSimpleOperationSequences2015, noiriFastQuantumInterface2018} that argued that the hybrid approach allowed them to utilize the respective strengths of the two qubit platforms. More specifically, the authors combine a singlet-triplet qubit with a Loss-DiVincenzo single-spin qubit. They take advantage of the significantly faster qubit initialization and readout of singlet-triplet qubits provided by the Pauli spin blockade (as opposed to the slow spin-selective tunneling to a lead used in the Loss-DiVincenzo qubit), and the exceptionally long coherence times of single-spin qubits (as opposed to the much shorter coherence times of singlet-triplet qubits). By contrast to these earlier studies, we do not focus on initialization or readout advantages of the specific qubit platform, but rather on the asymmetry that the heterogeneous encoding generates. Further, instead of using a different number of dots to construct the $A$ and $B$ qubit, our proposal uses the same number of dots for $A$ and $B$ making arbitrarily encoded arrays accessible on any singlet-triplet qubit device, including those that exist today.

\subsection{\label{sec:Out} Scaling Up}

As shown, CNOT or other asymmetric gates are best implemented in heterogeneous $AB$ qubit architectures.
Scaling up the two-qubit proposal presented here to multi-qubit devices is a natural direction for future research. To facilitate implementation of the CNOT gates between any two nearest neighbors, the multi-qubit device should continue the alternating $AB$ pattern. Although this goal can bring additional experimental challenges (such as gate cross-talk), these are the usual hurdles when the number of qubits on a semiconductor quantum dot device is increased. 

Although the proposed CNOT gate is, in principle, enough to achieve universal quantum computation when complemented with single-qubit gates, in practice certain computations will introduce significant gate overheads. For instance, here, we have directly addressed implementation of a CNOT gate between any two nearest-neighbor dots. However, if an operation on distant qubits is desired, it can be achieved either by qubit shuttling or by a consecutive application of the SWAP gates to bring them together and then return them back. This introduces  overhead since each SWAP requires 3 CNOT operations. Therefore, while it is not strictly necessary for universal computation, direct implementation of other two-qubit gates like SWAP could significantly reduce the gate overhead.

In contrast to CNOT, symmetric gates like SWAP benefit from the existing homogeneous qubit architecture. In practice,  the flexibility of the QD setup enables one to choose a qubit mapping that minimizes the number of elementary steps needed to implement a particular calculation, and this choice can be made immediately before the run, optimizing the qubit encoding for a particular quantum algorithm.

\subsection{Asymmetry as a Resource}
The central challenge in realizing high-fidelity entangling gates lies in the fundamental mismatch between the symmetric physical interactions of identical particles (e.g., Heisenberg exchange) and the asymmetric logical requirements of the CNOT gate. Traditionally, quantum information architectures resolve this mismatch at the ``software'' level via composite, multi-pulse sequences that artificially break the symmetry. Although functional, this approach inevitably increases logical gate depth, prolongs gate times, and exposes the system to further decoherence.

In an attempt to go beyond the limitations of the Heisenberg exchange interaction, various strategies have been proposed at the hardware level as well. One attempt~\cite{mehlTwoqubitCouplingsSinglettriplet2014} is to couple qubits indirectly via a mediator quantum state that interacts with both of them. 
In Ref.~\onlinecite{spethmannHighfidelityTwoqubitGates2024},  a superconducting qubit was considered as a mediator between two $ST_0$ qubits  and was shown to yield a longitudinal field Ising interaction between them, which is still symmetric and identical to the coupling between the resonant qubits. The Loss group also proposed mediating interactions between Majorana qubits using quantum-dot spin qubits, which enables a CNOT gate on Majorana qubits without interqubit braiding~\cite{hoffmanUniversalQuantumComputation2016}.
The other strategy to generate asymmetry utilizes spin-orbit coupling. Unlike electrons, holes in silicon are affected by strong spin-orbit interactions that create strong exchange anisotropy, which breaks the symmetry and facilitates the realization of asymmetric gates, such as the controlled rotation (CROT) gate~\cite{geyerAnisotropicExchangeInteraction2024}. We leave this promising direction for future research.

By contrast, our proposal demonstrates that this symmetry breaking can instead be offloaded directly into the physical hardware and interfaced with standard qubit designs in DQDs.  By alternating $ST_0$ and $ST_-$ encodings, the hardware naturally yields an asymmetric $XZ$ effective Hamiltonian. This physical asymmetry translates directly into a reduced quantum operation overhead, fulfilling the fundamental quantum information requirement for a fast, single-pulse control-target gate. Framing device engineering through this lens of exchange symmetry provides a clear theoretical framework for identifying and solving hardware bottlenecks at the physical level before they become algorithmic liabilities.

While we have demonstrated this principle in a DQD device, the underlying principle --- that a symmetric physical interaction acquires an asymmetric form when projected onto heterogeneously encoded logical subspaces --- is general. Therefore, other qubit platforms that feature identical physical constituents, symmetric two-body coupling, and the possibility of inequivalent qubit encodings within a single device can benefit from this. Trapped ions offer a particularly close analog. The M{\o}lmer–S{\o}rensen interaction~\cite{molmerMultiparticleEntanglementHot1999} generates an effective spin–spin coupling that is symmetric under exchange of the ions, while the \textit{omg} architecture~\cite{allcockOmgBlueprintTrapped2021} (employing optical-frequency, metastable-state and ground-state qubits) allows ground-state, metastable, and optical qubit encodings to coexist in a single ion species. Alternatively, a qubit may be encoded in the decoherence-free subspace $\{\ket{01},\ket{10}\}$ of an ion pair~\cite{kielpinskiDecoherencefreeQuantumMemory2001,lidarDecoherencefreeSubspacesQuantum1998} --- a direct analog of the $ST_0$ encoding in DQDs --- while a second is carried by a bare ion. In either case, the heterogeneous encoding could provide asymmetry for the interqubit interaction (see Sec.~\ref{sec:break}). Neutral-atom arrays provide a similar setting: the van der Waals interaction between Rydberg-excited atoms is  manifestly symmetric between identical atoms~\cite{jakschFastQuantumGates2000,saffmanQuantumInformationRydberg2010}, yet if one atom encodes its qubit in hyperfine or nuclear-spin ground states while its neighbor employs a ground–Rydberg or optical-clock encoding --- as realized in the $\ce{^{171}Yb}$ \textit{omg} architecture~\cite{chenAnalyzingRydbergbasedOpticalmetastableground2022,lisMidcircuitOperationsUsing2023} --- the projected interaction acts as a phase-type (Z) operator on one logical qubit and a population-dependent projector on the other, yielding precisely the encoding-induced asymmetry discussed here.

It is instructive to contrast this principle with superconducting circuits, where asymmetric effective interactions are well established but arise from a fundamentally different origin. In the cross-resonance gate~\cite{rigettiFullyMicrowavetunableUniversal2010}, a symmetric capacitive coupling between two fixed-frequency transmons yields an effective $ZX$ interaction when one qubit is driven at the frequency of the other; the asymmetry, however, is inherited from parameter inhomogeneity --- the deliberate detuning of two physically distinct circuits, which singles out control and target roles --- rather than from the structure of the qubit encoding. The same holds for hybrid transmon–fluxonium gates~\cite{cianiMicrowaveactivatedGatesFluxonium2022}, where the two circuits differ in their Hamiltonians from the outset and no symmetry exists to be broken. The closest superconducting analogue is therefore not found among circuit-QED two-qubit gates but in bosonic codes, where nominally identical cavities coupled by a symmetric beam-splitter or cross-Kerr interaction can host different logical encodings --- for example, a cat code in one mode~\cite{mirrahimiDynamicallyProtectedCatqubits2014} and a Gottesman–Kitaev–Preskill code~\cite{gottesmanEncodingQubitOscillator2001} in the other --- such that the projected logical interaction is again rendered asymmetric by the encoding alone. We therefore expect heterogeneous encoding to offer a hardware-efficient route to asymmetric effective interactions in any platform whose native couplings are symmetric, provided it features multiple inequivalent qubit encodings.

\section{\label{sec:Conclude}Conclusions}

We have identified a symmetry mismatch at the heart of two-qubit gate design: the interactions naturally available between identical particles---Heisenberg exchange and Coulomb repulsion---are invariant
under qubit exchange, while control--target gates such as CNOT are not. Whenever the qubits are encoded
identically (homogeneously), this symmetry must survive the projection onto the computational subspace. Homogeneous qubit encodings therefore cannot implement asymmetric control--target gates directly: the asymmetry must be supplied, for instance, by additional single- or two-qubit operations, at the cost of longer gate times and accumulated errors. 

By contrast, encoding neighboring qubits heterogeneously breaks the exchange symmetry at the hardware level, so that the same symmetric physical interaction projects onto an asymmetric effective interaction in the logical subspace. We have exploited this principle in semiconductor double quantum dots,
where multiple qubit encodings coexist on identical hardware and can be selected by gate voltages alone. Pairing an $ST_-$ with an $ST_0$ qubit converts the symmetric Heisenberg exchange into an asymmetric
$XZ$ coupling, enabling a single-pulse, all-electrical CNOT gate with a duration of $100$~ns. Our simulations account for the leakage out of the computational subspace, residual target-qubit splitting, and quasi-static charge noise as well as hyperfine noise and predict process fidelities above $99\%$ for parameters representative of isotopically purified silicon. This is achieved without microwave driving, strong spin--orbit interaction, or modifications to existing device layouts. The proposal is thus
directly testable on current singlet-triplet devices. 

More broadly, our results establish qubit encoding as a design resource in its own right: heterogeneous encoding converts a fixed, symmetric physical interaction into an asymmetric logical one, moving symmetry breaking from the pulse sequence into the hardware. Our finding has potential implications to any platform that combines identical physical constituents, symmetric two-body couplings, and inequivalent qubit encodings. 
\begin{acknowledgments}
This material is based upon work supported by the National Science Foundation under grant No. PHY-2310657. RK acknowledges helpful discussions with Anna Spak.
\end{acknowledgments}
\bibliography{Refs}

\end{document}